# Origin of multiple skyrmion phases in EuAl$_4$

Y. Arai[1], K. Nakayama[1,*], A. Honma[1], S. Souma[2,3], D. Shiga[4], H. Kumigashira[4], T. Takahashi[1], K. Segawa[5,*], and T. Sato[1,2,3,6,7,*]

[1]Department of Physics, Graduate School of Science, Tohoku University, Sendai 980-8578, Japan

[2]Advanced Institute for Materials Research (WPI-AIMR), Tohoku University, Sendai 980-8577, Japan

[3]Center for Science and Innovation in Spintronics (CSIS), Tohoku University, Sendai 980-8577, Japan

[4]Institute of Multidisciplinary Research for Advanced Materials (IMRAM), Tohoku University, Sendai 980-8577, Japan

[5]Department of Physics, Kyoto Sangyo University, Kyoto 603-8555, Japan

[6]International Center for Synchrotron Radiation Innov1ation Smart (SRIS), Tohoku University, Sendai 980-8577, Japan

[7]Mathematical Science Center for Co-creative Society (MathCCS), Tohoku University, Sendai 980-8578, Japan

*Corresponding authors. e-mail: k.nakayama@arpes.phys.tohoku.ac.jp (K.N.); segawa@cc.kyoto-su.ac.jp (K.S.); t- sato@arpes.phys.tohoku.ac.jp (T.S.)

**Abstract**

The Dzyaloshinskii-Moriya (DM) interaction has been considered essential for skyrmion formation, however, the discovery of skyrmion lattices (SkLs) in nominally centrosymmetric materials where the DM interaction is forbidden, such as Eu(Ga$_{1-x}$Al$_x$)$_4$, has challenged this established view. Recent structural investigations of Eu(Ga$_{1-x}$Al$_x$)$_4$ have further complicated this issue by revealing that the charge-density wave breaks local symmetry, theoretically allowing DM interaction. This raises a fundamental question: are the complex magnetic phases driven by the DM interaction or by alternative mechanisms? Here, using soft-x-ray angle-resolved photoemission spectroscopy, we determine the three-dimensional bulk electronic structure of Eu(Ga$_{1-x}$Al$_x$)$_4$, and elucidate the electronic origins of its rich magnetic orders. We directly observe an $x$-dependent Lifshitz transition leading to the emergence of a Fermi-surface pocket. Importantly, multiple nesting vectors derived from this pocket match the symmetries and periodicities of the multiple SkLs. Moreover, these nesting vectors can also account for other magnetic orders, such as the zero-field helical magnetism, suggesting a common electronic origin of the complex magnetic phases. These findings suggest that competing nesting-induced Ruderman-Kittel-Kasuya-Yosida interactions and their engineering can generate and control various SkLs and related topological spin textures.

**Introduction**

The field of topological materials science has witnessed a surge of interest in quasiparticles known as magnetic skyrmions (Fig. 1a), whose topological spin configurations generate emergent electromagnetic fields of the Berry curvature in real space and associated unconventional phenomena [1-6]. Conduction electrons interacting with these fields lead to, for instance, the topological Hall and Nernst effects [7-9]. Furthermore, skyrmions exhibit current-driven motion, such as the skyrmion Hall effect, even at ultralow current densities [10]. A key future direction involves shrinking of skyrmion size, which enhances both emergent field effects and potential application functionalities. In this context, the recent discovery of nanometric skyrmions (<5 nm) in centrosymmetric materials (e.g., $Gd_2PdSi_3$, $GdRu_2Si_2$, and $EuAl_4$ [11-14]), in contrast to larger skyrmions in conventional non-centrosymmetric materials (e.g., MnSi and FeGe [2,15]), is groundbreaking. However, the mechanism behind ultrasmall skyrmions in centrosymmetric materials remains unsettled.

In conventional non-centrosymmetric materials, a precursor helical spin structure (Fig. 1b) typically emerges from the Dzyaloshinskii-Moriya (DM) interaction with its strength $D$ (Fig. 1c) competing with ferromagnetic exchange coupling $J$. An applied field then transforms it to multi-**Q** helical spin states that are regarded as a moiré-like skyrmion lattice (SkL) [2,3,5,16]. The periodicity of SkL, i.e., skyrmion size, $\lambda = 2\pi/|\mathbf{Q}|$ is set by the ratio $Ja/D$ (where $a$ is the lattice constant) and typically ranges from 10 to 200 nm. The absence of DM interaction in centrosymmetric systems necessitates an entirely different mechanism for the formation of helical spin textures underlying a spin-moiré SkL [17-23]. One theoretical candidate for itinerant magnetic SkL system is the Ruderman-Kittel-Kasuya-Yosida (RKKY) interaction (Fig. 1d), wherein conduction electrons mediate the helical order of localized magnetic moments. This model suggests that the periodicity and orientation of the spin texture in the SkL phase are dictated by the Fermi-surface nesting vector **Q** [18,24-26], and allows for much smaller λ, potentially approaching the lattice constant (~1 nm). Remarkably, the RKKY framework also predicts that specific Fermi-surface topology can stabilize a rich variety of SkL phases and further complexed variants (e.g., anti-skyrmion, meron, and bimeron phases) [19,21,25,26]. These multi-**Q** states are energetically proximate to single-**Q** helical states. Disentangling a primary driving mechanism among multiple competing factors, such as magnetic frustration [11,17], many-body spin interaction [18,24,26], crystal/magnetic anisotropy [21], and magnetic dipole interaction [22], remains a key experimental challenge. Among these candidates, the RKKY interaction stands out as a particularly promising mechanism due to its direct connection with the Fermi-surface topology, making it a central focus of this study. For this sake, exploring a role of the RKKY interaction, especially the link between electronic structure and magnetic complexity, demands clarifying the bulk electronic structure and correlating it directly with tunable spin textures in real space.

EuAl$_4$ and its Ga-substituted counterparts, Eu(Ga$_{1-x}$Al$_x$)$_4$ (Fig. 1e, f) [27-29], were proposed to be an ideal platform to address this challenge because of their nominally centrosymmetric structure at room temperature. Elemental substitution drives the system through a quantum phase transition, transforming the zero-field ground state from the single-**Q** helical order in EuAl$_4$ ($x = 1$; the Néel temperature $T_N = 15.4$ K; see Fig. 1f) [13,30-33] to the A-type antiferromagnetism in EuGa$_4$ ($x = 0$; $T_N = 16.5$ K; Fig. 1e) [31,34] across the critical composition range of $0 < x_c < 0.4$ [35-37]. Significantly, under applied fields, EuAl$_4$ itself exhibits two distinct multi-**Q** SkL phases, i.e., a lower-field rhombic SkL (Fig. 1g) and a higher-field square SkL (Fig. 1h), adjacent to additional single/double-**Q** spin states [13]. This rich complexity, absent in previously studied centrosymmetric single-SkL hosts like Gd$_2$PdSi$_3$ and GdRu$_2$Si$_2$ [11,12], provides a valuable and rigorous testbed for the RKKY model. However, previous electronic-structure studies have focused primarily on the simpler single-SkL systems [38-42]. In EuAl$_4$, recent works provided a crucial theoretical framework that degenerate Fermi-surface nesting channels and associated RKKY interactions are key to stabilizing the magnetic phases in EuAl$_4$ [13,33,43], and demonstrated qualitative experimental support of this mechanism for the helical magnetism [33]. However, the experimental link between the underlying electronic structure and the emergence of multiple, symmetry-distinct SkL phases remains largely unexplored. Furthermore, very recent high-resolution diffraction studies have revealed that charge-density wave (CDW) [27,28,30,33,35,36,44], which coexists with magnetism below ~140 K, breaks inversion symmetry [45,46] and competes with the helical magnetism [46,47]. This discovery implies that the DM interaction is symmetry-allowed in the magnetic phase, challenging the previous assumption that these systems are purely governed by centrosymmetric physics. Disentangling the interplay between the DM and RKKY mechanisms is crucial for identifying the primary driving force that stabilizes diverse topological spin textures.

Here, we address this issue by performing a comprehensive study of the electronic state in Eu(Ga$_{1-x}$Al$_x$)$_4$ ($x$ = 0, 0.38, 0.5, and 1.0) using angle-resolved photoemission spectroscopy (ARPES) with energy-tunable soft-x-ray (SX) photons. This technique overcomes the limitation of surface sensitivity and significant $k_z$ broadening effects inherent to conventional ARPES, and provides access to the intrinsic three-dimensional (3D) bulk electronic structure, which allows us to move from a qualitative to quantitative understanding about the relationship between the electronic structure and bulk magnetisms. Furthermore, our SX-ARPES measurements reveal a Lifshitz transition across the critical composition range where the magnetic quantum phase transition occurs. We demonstrate that the observed change in the electronic structure governs the Fermi-surface nesting condition, and discuss its significance for enhancing the nesting-driven RKKY interaction to promote the helical spin texture and ultrasmall SkL phases.

**Results**

**Electronic structure of EuAl$_4$**

We first investigated the electronic structure of an end member EuAl$_4$ ($x$ = 1), in its paramagnetic phase ($T$ = 30 K). To elucidate the intrinsic bulk electronic states, we performed SX-ARPES measurements throughout the 3D Brillouin zone (Fig. 2a). Photon-energy ($h\nu$)-dependent measurements to probe the band dispersion along the surface-normal direction ($k_z$) reveal a key feature near the Fermi level ($E_F$), notably, an electron-like band (labeled e1) centered at the Z point. Nearly flat bands observed around the binding energy of 1.5 eV correspond to localized Eu 4$f$ states. Our density functional theory (DFT) calculations (red curves in Fig. 2b, excluding Eu 4$f$ states) reasonably reproduce the e1 band. The Fermi-surface map in the $k_x$-$k_z$ plane (Fig. 2c) shows that the e1 band forms a smaller elliptical pocket elongated along $k_z$ near the Z point (dashed green curve). In addition, there are a larger ellipse near the Z point and an open sheet with pronounced warping near the Brillouin zone boundary (dashed blue curves). As demonstrated in the next paragraph, both are formed by the same band, labelled e2. The observed clear $k_z$ dependence of both e1 and e2 confirms their bulk nature.

To gain further insight into the Fermi-surface topology, particularly its relevance for nesting, we performed detailed measurements for the band dispersion (Fig. 2d) and Fermi surface (Fig. 2e) at the Z (light blue shade in Fig. 2a) and Γ (light green shade) planes. These were simultaneously accessed using $hv$ = 424 eV (see the red line in Fig. 2c for the corresponding $k_z$ values). At the Γ plane, the Fermi surface exhibits a diamond-shaped feature centered at the Γ point (dashed blue curve in Fig. 2e), corresponding to the neck of the warped e2 Fermi surface seen in Fig. 2c (light blue arrow; see also Fig. 2d, where one can find an electron-like e2 band, which intersects with a hole-like band near $E_F$ according to the DFT calculations). The Z plane shows a more complex structure (Fig. 2d, e), which is crucial for the magnetism as discussed later. First, there is a small electron-like Fermi surface centered at the Z point, originating from the e1 pocket (dashed green curve). Second, petal-shaped electron-like pockets formed by e2 appear between Z and Σ (dashed blue curve). The portion of the e2 Fermi surface closer to the Z point ($k_x$ ~ ±0.3 π/a) corresponds to the larger elliptical Fermi surface in Fig. 2c (red arrows), whereas that closer to the Σ point ($k_x$ ~ ±0.8 π/a) corresponds to the maximum of the warped Fermi surface in Fig. 2c (purple arrows). Third, a large star-shaped Fermi surface (h1; dashed red curve) is present, though its band dispersion is less prominent as seen in Fig. 2c, d due to the weak intensity along the $k_x$ axis. The e1 and e2 pockets primarily originate from the Eu 5$d$ and Al 3$p$ orbitals, while the h1 pocket is dominated by the Al 3$p$ orbitals (Supplementary note 1). Overall, the experimental Fermi surface of EuAl$_4$ consists of three key elements: the inner 3D e1 Fermi surface, the middle 3D e2 Fermi surface, and the outer quasi-two-dimensional (quasi-2D) h1 Fermi surface. These are qualitatively captured by DFT calculations, as shown in the 3D Fermi-surface plot in Fig. 2f, while the calculated Fermi surface is slightly more complex due to hybridization effects.

**Lifshitz transition induced by elemental substitution**

Having established the electronic structure of EuAl$_4$ ($x = 1$), we next investigate its evolution upon Ga substitution in Eu(Ga$_{1-x}$Al$_x$)$_4$. A side-by-side comparison of band dispersions for $x = 0$, 0.5, and 1 (Fig. 3a–c) reveals a striking change near the Z point: notably, the e1 band clearly crosses $E_F$ at $x = 0.5$ and 1.0 but is absent at $x = 0$. This disappearance reflects an upward energy shift of the e1 band as $x$ decreases below 0.5. This trend is qualitatively reproduced by DFT calculations (Fig. 3d–f) and further corroborated by the absence of ARPES intensity near the Z point in the Fermi-surface map (Fig. 3g). These observations provide definitive evidence for the Lifshitz transition, i.e., a change in the Fermi-surface topology driven by chemical substitution. Additional measurements at $x = 0.38$ (Supplementary note 2) reveal the $E_F$ crossing of the e1 pocket, indicating that the transition occurs between $x = 0$ and $x = 0.38$. In addition to this topology change, the electronic structure further exhibits a quantitative evolution with $x$, such as a non-monotonic variation of the Dirac/Weyl nodal ring in the Z plane [48] (Supplementary note 3).

**Nesting-driven helical magnetism triggered by Lifshitz transition**

Intriguingly, the critical composition range for the observed Lifshitz transition ($0 < x < 0.38$) coincides with the onset of helical magnetism at zero magnetic field [13,30,35-37]. This remarkable correlation strongly suggests a close link between the emergent e1 pocket and the helical order. To test this hypothesis within the RKKY interaction, where the Fermi-surface nesting dictates the magnetic ordering vector **Q**, we quantitatively analyzed the nesting conditions (Fig. 4). We extracted the experimental Fermi wave vectors in the $k_x$-$k_y$ ($k_z = \pi$) and $k_x$-$k_z$ ($k_y = 0$) planes for different $x$ (Fig. 4a, d–f and Fig. 4b, g–i, respectively; see Supplementary note 4 for the ARPES intensity in the $k_x$-$k_z$ plane for $x = 0$, 0.38, and 0.5). Quasi-straight, parallel Fermi-surface segments, generally contributing to nesting, are highlighted by filled circles. For $x = 1$ (EuAl$_4$), such segments exist on the e1 Fermi surface along both $k_x$ and $k_y$ (see green circles in Fig. 4a) as well as $k_z$ (Fig. 4b). The inner parts of the e2 pockets also show quasi-straight segments (blue circles). These features facilitate nesting along the $k_x$ or $k_y$ directions, in consistent with the observed in-plane nature of the helical magnetism. Comparing the potential nesting vectors **Q**$_{e1-e1}$, **Q**$_{e1-e2}$, and **Q**$_{e2-e2}$ connecting quasi-straight segments, we find that **Q**$_{e1-e2}$, which connects the e1 and e2 pockets (red arrow in Fig. 4a, b), quantitatively matches the experimental helical wave vector **Q**$_1 \sim (0.19, 0, 0)$ (compare red circles and dashed red line in Fig. 4c). This agreement persists for $x = 0.38$ and 0.5 (Fig. 4k, l), where the helical order exists, and vanishes at $x = 0$ (Fig. 4j), where the e1 pocket is absent and helical order gives way to A-type antiferromagnetism. This provides compelling evidence that the zero-field helical magnetism is driven by nesting-induced RKKY interactions involving the e1 and e2 bands.

**Discussion**

**Mechanisms stabilizing multiple SkLs**

We now discuss the origin of the most remarkable characteristic of EuAl$_4$: the multistep field-induced transitions involving two distinct SkLs (Fig. 1g, h). This complexity offers a unique opportunity to test the RKKY mechanism. We first examine the square SkL, which has a simpler structure. This SkL is characterized by the helical ordering vectors $Q_4$ // [110] and $Q_5$ // [1$\bar{1}$0] (Fig. 1h), rotated by 45° relative to the zero-field helical $Q_1$ // [100] (Fig. 1f). By shifting the measured Fermi surface under zero magnetic field by $Q_4$ (red lines in Fig. 4n), we observe overlap with the original Fermi surface (blue lines), specifically an inter-band nesting between the e1 and e2 pockets. The nested quasi-straight segments are highlighted in purple (also indicated by yellow arrows). Quantitative analysis confirms that the nesting vector along the [110] direction (green circles in Fig. 4o) well matches the magnetic ordering vector $Q_4$ (dashed green line). This key finding indicates that the nesting-driven RKKY interaction also stabilizes the helical spin texture with $Q_4$, which is distinct from the zero-field helical magnetism with $Q_1$ (see Fig. 4m for comparison of $Q_1$ and $Q_4$ in $q$ space).

Due to the $C_4$ symmetry of the underlying Fermi surface, nesting simultaneously favors equivalent interactions at the symmetry-related vectors $Q_1$' and $Q_4$' (rotated 90° from $Q_1$ and $Q_4$, respectively) (Fig. 4m). Therefore, our ARPES results reveal the presence of at least four competing nesting instabilities ($Q_1$, $Q_1$', $Q_4$, $Q_4$') in EuAl$_4$. This contrasts with previously studied centrosymmetric skyrmion-hosts such as Gd$_2$PdSi$_3$ and GdRu$_2$Si$_2$ that would host only two non-equivalent vectors, and would explain the field-induced $Q$ reorientation driven by magnetic frustration. Furthermore, the superposition of helical spin textures with $Q_4$ and $Q_4$' naturally forms the square SkL (compare Figs. 1h and 4m). The energy gain for this multi-$Q$ state may originate from high-harmonic wave-vector interactions [26,43,49] at $Q_1$ and $Q_1$', which correspond to the second harmonics of $Q_4$ and $Q_4$' ($Q_1 = Q_4 + Q_4$' and $Q_1$' $= Q_4 − Q_4$').

The next key question is the formation mechanism of rhombic SkL. This phase is not only intriguing in the context of multiple SkLs but also due to its rotational symmetry breaking, reminiscent of nematicity in strongly correlated systems. Namely, while SkLs usually preserve the lattice symmetry of the paramagnetic phase, the rhombic SkL in EuAl$_4$ exhibits C$_2$ symmetry, breaking the original tetragonal C$_4$ symmetry. This is partly related to spin-lattice coupling and a subtle orthorhombic lattice distortion along [100] [44] (note that this distortion is suppressed in the square SkL phase). From an electronic-structure point of view, such a distortion would lift the degeneracy between nesting involving $\mathbf{Q}_1$ and $\mathbf{Q}_1$' while preserving it for $\mathbf{Q}_4$ and $\mathbf{Q}_4$'. This electronic anisotropy could favor a triple-Q state consisting of, for instance, $\mathbf{Q}_1$, $\mathbf{Q}_4$, and $\mathbf{Q}_4$' (effectively removing $\mathbf{Q}_1$'-related components from Fig. 4m), yielding a C$_2$ symmetric magnetic structure consistent with observations in the rhombic SkL phase (Fig. 1g) [43]. The expected electronic anisotropy needs to be clarified by direct low-temperature measurements in the magnetic phase in future.

To the best of our knowledge, SkLs have been directly observed only around $x = 1$ in Eu(Ga$_{1-x}$Al$_x$)$_4$. Their evolution as a function of composition is an unresolved key issue. To discuss an implication of the present results in relation to this issue, we have performed the analysis of the $\mathbf{Q}_4$ vector in the doped samples ($x = 0.38$ and 0.5) and found that the nesting channel $\mathbf{Q}_{e1-e2}$ along the [110] direction persists in these compositions (Supplementary note 5). We estimate the $\mathbf{Q}_4$ vectors from the average of $\mathbf{Q}_{e1-e2}$ to be (0.089, 0.089, 0) for $x = 0.38$ and (0.072, 0.072, 0) for $x = 0.5$, which are similar to $\mathbf{Q}_4 =$ (0.083, 0.083, 0) determined by scattering measurements for $x = 1$ [13]. This finding provides a testable prediction: if SkLs also exist in these doped compounds, their periodicities would only weakly depend on composition.

We now comment on the role of the CDW order reported in Eu(Ga$_{1-x}$Al$_x$)$_4$ [27,28,30,33,35,36,44]. The CDW order in Eu(Ga$_{1-x}$Al$_x$)$_4$ has out-of-plane **Q** vector [e.g., **Q**$_{CDW}$ = (0, 0, 0.183) at $x$ = 1 [30,33]], breaks inversion symmetry, and may influence the spin order [45-47]. Since our ARPES measurements were performed at $T$ = 30 K, below the CDW transition temperatures of ~140 K for EuAl$_4$ ($x$ = 1.0) and ~50 K for EuGa$_2$Al$_2$ ($x$ = 0.5), the band structure already includes the influence of CDW. Consequently, the similarity of the observed band structure with that predicted by the DFT calculations in the absence of CDW suggests that the characteristic Fermi-surface topology driving the RKKY interaction remains robust even within the CDW phase, particularly at the $k_z = \pi$ plane, to which we mainly pay attention. In addition, the in-plane **Q**$_1$ and **Q**$_4$ nesting vectors favor instabilities toward the helical magnetic and SkL phases rather than the out-of-plane CDW. Therefore, although the non-centrosymmetric structure of the CDW phase in Eu(Ga$_{1-x}$Al$_x$)$_4$ allows DM interaction, RKKY interaction plays a dominant role in determining the magnetic modulations. This further suggests a need to re-evaluate the relative importance of itinerant frustration versus DM interaction in a broader range of non-centrosymmetric magnets where itinerant interactions like RKKY may have been overlooked.

**Unified picture of complex magnetism**

Our identification of four competing nesting channels, all of which critically depend on the existence of the e1 pocket formed through the Lifshitz transition, provides a unified framework for understanding the exceptionally rich magnetic phase diagram of EuAl$_4$. Beyond the helical (called Phase I in the original work), rhombic SkL (Phase II), and square SkL (Phase III), EuAl$_4$ hosts a variety of exotic states, including vortex-antivortex lattice at high fields and low temperatures (Phase IV), a spiral screw state (Phase V) and a meron-antimeron lattice at and near zero magnetic field above the helical magnetic phase and below $T_N$ (Phase VI) [13]. Remarkably, combination and selection of **Q**$_1$, **Q**$_1$', **Q**$_4$, and **Q**$_4$' can construct the magnetic structures of all these phases [13,43]. This demonstrates that the intricate interplay of competing RKKY interactions, governed by the Fermi-surface topology revealed by our SX-ARPES measurements, drives the complex landscape of topological and exotic magnetic phases and their tunability.

**Methods**

**Sample fabrication**

Growth of the single crystals was carried out in the same way as described in the literature [28]. However, for the intermediate phase of Eu(Ga$_{1-x}$Al$_x$)$_4$ except EuGa$_4$ and EuAl$_4$, we found that flux removal could be achieved by soaking in alcohol, rather than centrifugation. The actual compositions of the single crystals were measured using ICP-AES (inductively coupled plasma atomic emission spectroscopy) analysis.

**Calculation**

First-principles band-structure calculations were conducted using Quantum ESPRESSO code [50,51] within the generalized gradient approximation (GGA) scheme using PSLibrary [52]. The experimental lattice parameters and the Wyckoff positions determined by x-ray diffraction and x-ray absorption fine structure measurements [28], respectively, were used. The spin-orbit coupling was included consistently in calculating electronic structures with uniform $k$-point mesh 8×8×8. The crystal structure and the Fermi surfaces have been visualized with VESTA [53].

**ARPES measurements**

SX-ARPES measurements were conducted with an Omicron-Scienta SES2002 electron analyzer with energy-tunable synchrotron light at BL-2A in Photon Factory (PF), KEK. We used linearly polarized light (horizontal polarization) of 280–580 eV, with an energy resolution of around 50 meV. All ARPES measurements were conducted in the paramagnetic phase. Samples were cleaved in situ along the (001) plane and measured under an ultrahigh vacuum better than $1\times10^{-10}$ Torr. The crystal orientation was determined by x-ray backscattering measurement, which confirmed the apparent four-fold rotational symmetry of the samples, consistent with the (001) cleaved plane.

**Data availability**

The data that support the findings of this study are available within the main text and Supplementary Figures. Any other relevant data is available from the corresponding authors upon request.

**Acknowledgments**

We thank Y. Onose and H. Masuda for fruitful discussion. This work was supported by JST-CREST (no. JPMJCR18T1 to T.S.), Grant-in-Aid for Scientific Research (JSPS KAKENHI Grant Number JP21H04435 to T.S., JP23H01115 to K.N.), KEK-PF (Proposal No. 2024S2-001, 2022G652, 2024G141, 2024G136). Y. A. and A.H. acknowledge support from GP-Spin at Tohoku University. A.H. also thanks JSPS.

**Author contributions**

The work was planned and proceeded by discussion among Y.A., K.N., K.S., and T.S. K.S. grew single crystals. Y.A., K.N., A.H., S.S., D.S., H.K., T.T., and T.S. performed the ARPES measurements. Y.A. performed band structure calculations. Y.A. and K.N. wrote the manuscript with inputs from all the authors.

**Competing Interests**

The authors declare no competing interests.

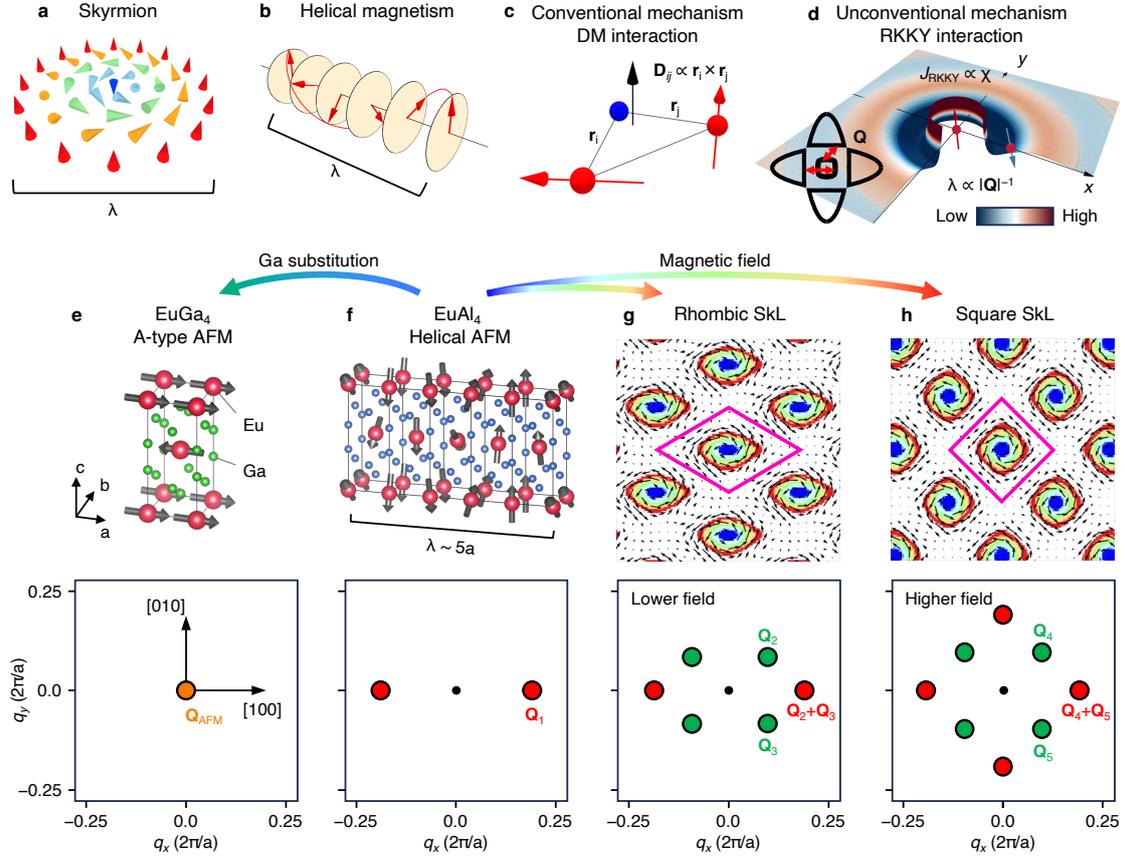

**Fig. 1 | Skyrmion formation mechanism and magnetic structures of Eu(Ga$_{1-x}$Al$_x$)$_4$. a, b** Schematics of a two-dimensional Bloch-type skyrmion and a one-dimensional helical magnetic order, respectively. **c, d** Schematics of Dzyaloshinskii-Moriya (DM) and Ruderman-Kittel-Kasuya-Yosida (RKKY) interactions, respectively, both of which are potential mechanisms for helical magnetism and skyrmion formation. The DM interaction, described by $\mathbf{D}_{ij} \propto \mathbf{r}_i \times \mathbf{r}_j$, favors twisted alignment of adjacent magnetic moments in non-centrosymmetric systems. The RKKY interaction ($J_{RKKY}$) induces long-range ordering of localized magnetic moments mediated by itinerant carriers, with the period ($\lambda$) inversely proportional to the nesting vector **Q**. **e** Schematic of crystal structure and A-type commensurate antiferromagnetic (AFM) order in EuGa$_4$ (top), characterized by ferromagnetic spin arrangements within Eu layers and antiferromagnetic coupling

between layers. The corresponding magnetic wave vector $\mathbf{Q}_{AFM}$ = (0, 0, 0) (bottom) reflects the fact that the nonmagnetic unit cell contains two Eu layers. **f** Same as **e**, but for EuAl$_4$, which exhibits incommensurate helical AFM order at zero field (top), characterized by spin rotation along the *a* axis with $\mathbf{Q}_1$ ~ (0.19, 0, 0) (bottom). **g, h** Schematics of field-induced rhombic and square skyrmion-lattice (SkL) states with exceptionally small diameters (top), characterized by multi-$\mathbf{Q}$ structures (bottom): $\mathbf{Q}_2$ ~ (0.097, 0.073, 0), $\mathbf{Q}_3$ ~ (0.097, -0.073, 0), $\mathbf{Q}_4$ ~ (0.083, 0.083, 0), and $\mathbf{Q}_5$ ~ (0.083, -0.083, 0) [13].

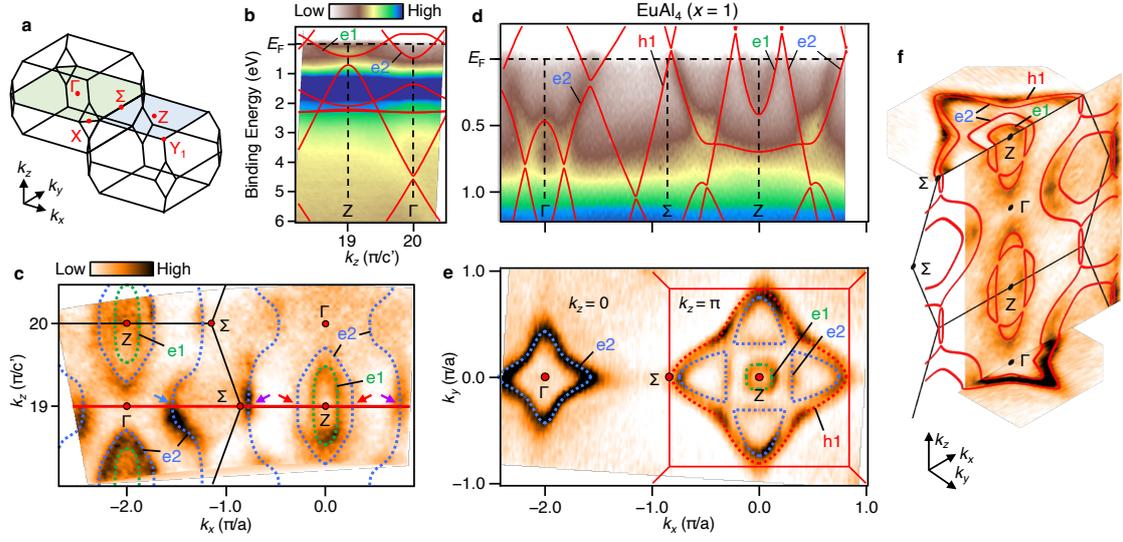

**Fig. 2 | Band structure of EuAl$_4$ ($x$ = 1) in paramagnetic state. a** Brillouin zone of EuAl$_4$. **b** Angle-resolved photoemission spectroscopy (ARPES) intensity at 30 K obtained in the normal-emission set up with varying photon energy ($h\nu$) from 390 eV to 490 eV, plotted as a function of binding energy and $k_z$ ($c$' is defined as 0.5$c$ due to the periodicity of primitive unit cell). The $k_z$ values are calculated from $h\nu$ using the formula $k_z = \sqrt{2m_e(h\nu - \phi - E_B + V_0)}/\hbar - (\omega \cos\eta)/c$, where $m_e$ is the electron mass, $\phi$ is the work function of sample, $V_0$ is the inner potential (set to 16 eV), and $\eta$ is the angle between the incident light and the normal of the sample. Red curves indicate the bulk band dispersions along the Γ-Z line, calculated using the density functional theory (DFT). **c** ARPES intensity at the Fermi energy ($E_F$) with an integrated energy window of ±40 meV, plotted as a function of $k_z$ and $k_x$. Dashed green and blue curves are a guide for the eyes to trace the experimental Fermi wave vectors ($k_F$'s) of the e1 and e2 pockets, respectively. **d** Plot of near-$E_F$ band structure along the Γ-Σ-Z path (red line in **c**), probed by 424-eV photons, at $T$ = 30 K. The corresponding DFT calculations are overlaid in red curves. **e** Plot of ARPES intensity at $E_F$ as a function of $k_y$ and $k_x$ (around the Γ-Σ-Z line), obtained by 424-eV photons. Dashed green, blue, and red curves are a guide for the eyes

to trace the e1, e2, and h1 Fermi surfaces, respectively. **f** Three-dimensional Fermi surface constructed by combining the data from **c** and **e**, together with calculated Fermi surface (red curves).

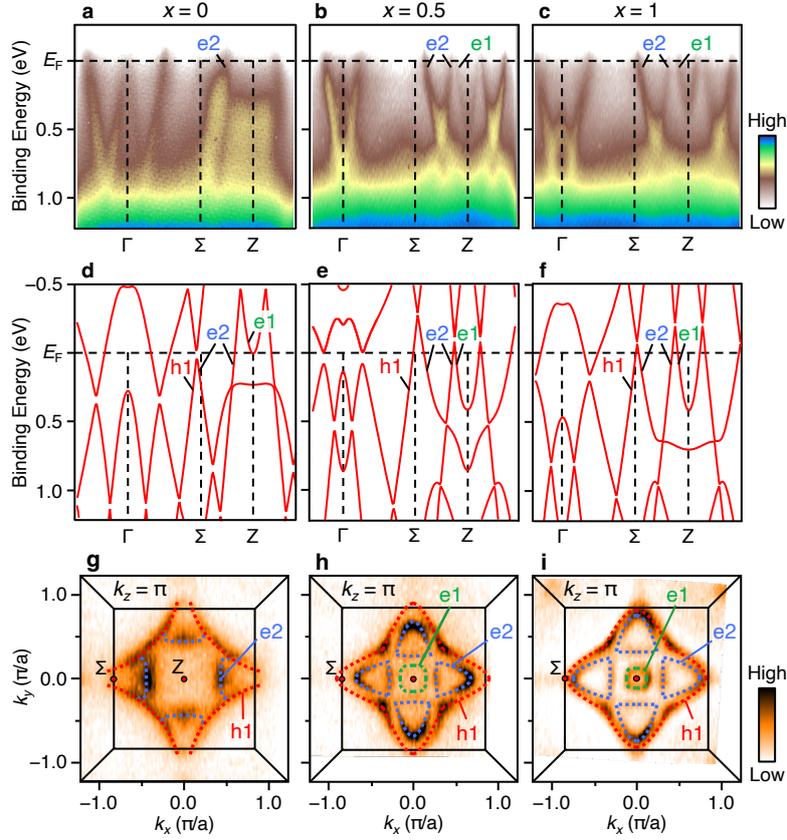

**Fig. 3 | Observation of Lifshitz transition induced by elemental substitution. a–c** Evolution of the paramagnetic band structure along the Γ-Σ-Z line for $x = 0$, 0.5, and 1.0. Presence of e1 pocket at the Z point for $x = 0.5$ and 1.0, and its absence for $x = 0$, indicates Lifshitz transition. **d–f** Calculated band structures along the Γ-Σ-Z line for $x = 0$, 0.5, and 1.0. **g–i** Compositional dependence of Fermi surfaces in the $k_z = \pi$ plane, supporting the occurrence of Lifshitz transition.

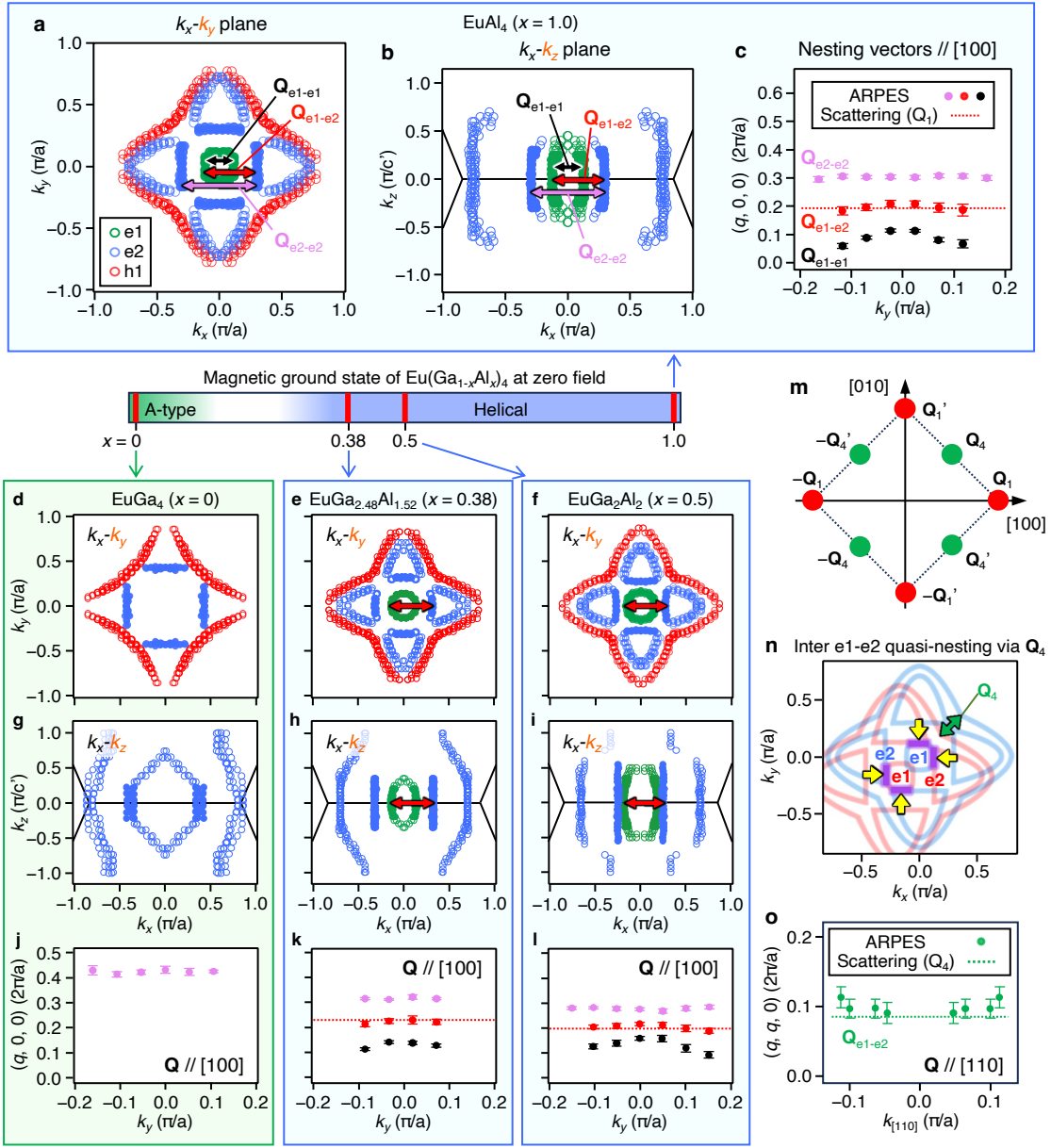

**Fig. 4 | Relationship among Fermi-surface topology, helical magnetism, and skyrmion lattice.** **a** Experimental $k_F$ values in the Z plane for $x = 1.0$ under zero magnetic field, deduced from Fig. 3i. Quasi-straight Fermi-surface segments are shown by filled circles. Nesting vectors between the straight segments of the e1 pocket ($\mathbf{Q}_{e1-e1}$), the e1 and e2 pockets ($\mathbf{Q}_{e1-e2}$), and the e2 pocket ($\mathbf{Q}_{e2-e2}$) are indicated by black, red, and magenta arrows, respectively. **b** Experimental $k_F$ values in the $k_x$-$k_z$ plane for $x = 1.0$, extracted from Fig. 2c. **c** Plot of $\mathbf{Q}_{e1-e1}$, $\mathbf{Q}_{e1-e2}$, and $\mathbf{Q}_{e2-e2}$ as a function of $k_y$, estimated from **a** and

**b**. The $|Q_1|$ value determined by neutron scattering measurements [13] is plotted by a dashed red line. The error bars correspond to the standard error of the fit. **d–f, g–i**, and **j–l** Same as **a**, **b**, and **c**, respectively, but for $x = 0$, 0.38, and 0.5 (deduced from Figs. 3. a, b, g, h, and S2). **m** Schematic illustration of experimentally determined $Q_2$ and $Q_4$ vectors of EuAl$_4$ in $q$ space [13]. **n** Comparison of Fermi surfaces between original (blue) and shifted by $Q_4$ (red) in EuAl$_4$. Nested segments are indicated by yellow arrows. **o** Plot of nesting vector along [110] estimated from **a**. Dashed green line corresponds to $Q_4$ reported by scattering measurements [13].